\documentclass[a4paper,11pt]{article}
\usepackage{pos}
\usepackage{orcidlink}
\usepackage{subfig}
\usepackage{diagbox}
\usepackage{makecell, multirow}
\usepackage{pgf}
\usepackage{soul}
\usepackage{graphicx, color}
\usepackage{rotating}
\usepackage{physics}
\usepackage{booktabs}
\allowdisplaybreaks

\definecolor{link_blue}{RGB}{51,102,204}
\hypersetup{%
pdftitle = {title},
pdfsubject = {},
pdfkeywords = {},
colorlinks = {true},
filecolor = {black},
linkcolor = {link_blue},
menucolor = {black},
citecolor = {link_blue},
urlcolor = {link_blue},
}{}

\title{A novel Hamiltonian formulation of $1+1$ dimensional $\phi^4$ theory in Daubechies wavelet basis: momentum space analysis}
\ShortTitle{A novel Hamiltonian formulation of $1+1$ dimensional $\phi^4$ theory in Daubechies wavelet basis: momentum space analysis}

\author*[a]{Mrinmoy Basak\orcidlink{0000-0001-8965-0824}}

\affiliation[a]{Department of Theoretical Physics, Tata Institute of Fundamental Research,\\Homi Bhabha Road, Mumbai 400005, India}

\emailAdd{mrinmoy.263009@gmail.com}
\abstract{We employ the wavelet formalism of quantum field theory to study field theories in the nonperturbative Hamiltonian framework. Specifically, we make use of Daubechies wavelets in momentum space. These basis elements are characterised by a resolution and a translation index that provides for a natural nonperturbative infrared and ultraviolet truncation of the quantum field theory. As an application, we consider the $\phi^4$ theory and demonstrate the emergence of the well-known nonperturbative strong-coupling phase transition in the $m^2>0$ sector.}

\FullConference{The 42nd International Symposium on Lattice Field Theory (LATTICE2025)\\
2-8 November 2025\\
Tata Institute of Fundamental Research, Mumbai, India\\}

\begin{document}
\maketitle
\section{\label{sec:Introduction}Introduction}

Although the Hamiltonian formalism is conceptually fundamental, it was largely de-emphasized in relativistic quantum field theory (QFT) during the 1950s and 1960s due to the success of covariant perturbative methods \cite{PhysRev.76.769} and the lack of computational resources for nonperturbative diagonalization. However, between 1960 and 1980, interest in the Hamiltonian framework revived \cite{PhysRev.140.B445,PhysRevD.2.1438, PhysRevD.10.2445, PhysRevD.11.395, PhysRevD.19.3715, PhysRev.125.397, PhysRev.128.2425, COLEMAN1975267, COLEMAN1976239, PhysRevD.11.2088}, driven by its ability to access the full spectrum and nonperturbative structure that perturbation theory obscures. This revival gained momentum with Wilson’s development of the renormalization group (RG) \cite{PhysRevB.4.3174} and has accelerated in the modern era with improved numerical techniques like DMRG and tensor networks. Today's age of quantum simulations, Hamiltonian methods are viewed as a primary avenue for \textit{ab initio} computation of real-time dynamics, particularly where Euclidean lattice Monte Carlo fails \cite{PhysRevD.110.096016, PRXQuantum.3.020324, doi:10.1098/rsta.2021.0069, doi:10.1098/rsta.2021.0062, PhysRevA.105.032418}.

In 1965, Wilson introduced "wave-packet" basis functions as an alternative to the standard Fourier basis for resolving quantum fields \cite{PhysRev.140.B445}. He sought basis elements simultaneously localized in both momentum and position space to solve the Hamiltonian eigenvalue problem beyond perturbation theory. While Wilson constructed functions that approximated this behavior, the rigorous mathematical realization of such compactly supported, orthonormal bases—now known as wavelets—was achieved decades later. The concept was formalized in the 1980s by Morlet and Grossmann \cite{doi:10.1190/1.1441328,doi:10.1190/1.1441329,doi:10.1137/0515056}, and fully matured with the development of multiresolution analysis (MRA) and compactly supported wavelets by Meyer, Mallat, and Daubechies \cite{Meyer1986-1987, Mallat1989MultiresolutionAA, daubechies1992ten}.

Following these mathematical advances, Wilson re-emphasized the potential of wavelets for nonperturbative QCD in 1994 \cite{PhysRevD.49.6720}, specifically advocating for their use in similarity renormalization group (SRG) procedures to achieve scale-separated, band-diagonal effective Hamiltonians. While wavelet techniques have since been applied to regularization and gauge theories in various contexts \cite{kessler2003waveletnotes, 10.1007/s00601-018-1357-z, michlin2017using, federbush1995new, BEST2000848, PhysRevLett.116.140403, 10.1007/JHEP06(2021)077, PhysRevD.106.036025, PhysRevD.108.125008, best1994variationaldescriptionstatisticalfield, HALLIDAY1995414, 10.1063/1.1543582, 10.1088/1751-8121/ad5503, PhysRevD.87.116011, 10.1051/epjconf/201817511002, 10.3842/SIGMA.2007.105, Albeverio2009ARO, 10.1134/S1063778818060029, PhysRevD.88.025015, 10.1007/s11182-013-9940-8, 10.1007/s10773-015-2913-7, polyzou2020lightfrontquantummechanicsquantum, PhysRevD.107.036015, PhysRevD.111.096024}, the literature has predominantly focused on position-space formulations (e.g., \cite{PhysRevD.87.116011}). The original proposal to analyze QFTs using a \textit{momentum-space} wavelet basis remains largely unexplored.

In this work, we address this gap by formulating the $1+1$ dimensional $\phi^4$ theory using a Daubechies wavelet basis in momentum space. We expand the creation and annihilation operators in terms of wavelet modes characterized by resolution ($k$) and translation ($n$) indices. The compact support of the Daubechies basis ensures that the significant contributions arising from a limited number of degrees of freedom. Unlike standard Hamiltonian truncation schemes that use free-field energy cutoffs \cite{PhysRevD.91.085011, PhysRevD.93.065014, 10.21468/SciPostPhys.13.2.011, 10.1007/JHEP10(2016)050, 10.1007/JHEP05(2021)190}, we construct the interacting Hamiltonian within a finite Fock space spanned by wavelet modes, compute the energy spectrum and demonstrate that this formalism can reliably track the strong-coupling phase transition in the $m^2 > 0$ sector.

The paper is organized as follows: Section \ref{sec:Daubechies_wavelet_basis} reviews the construction of the Daubechies wavelet basis. Section \ref{sec:formalism} details the wavelet-based Fock space and the free scalar field spectrum. Section \ref{sec:phi4_hamiltonian} presents the interacting $\phi^4$ Hamiltonian, the matrix element construction, and the analysis of the symmetry-breaking phase transition. Conclusions and future outlooks are provided in Section \ref{sec:conclusion}.
\section{\label{sec:Daubechies_wavelet_basis}Daubechies wavelet basis}

In this section, we outline the Daubechies wavelet basis formalism used in this work. For a comprehensive review, we refer the reader to \cite{PhysRevD.95.094501,kessler2003waveletnotes,PhysRevD.87.116011,https://doi.org/10.1002/cpa.3160410705,daubechies1992ten,PhysRevD.107.036015,388960,PhysRevD.111.096024}.

The Daubechies basis constitutes an orthonormal basis for $\mathcal{L}^2(\mathbb{R})$ composed of scaling functions and wavelet functions. The mother scaling function, $s(x)$, is defined by the renormalization group equation, also known as the scaling equation:
\begin{equation}
\label{eq:scaling_equation}
s(x) = \sum_{l=0}^{2K-1} h_l \sqrt{2} s(2x - l).
\end{equation}
Here, $K$ is the order of the wavelet (determining the support $[0, 2K-1]$) and $h_l$ are the filter coefficients. We utilize the translation operator $\hat{T}f(x) = f(x-1)$ and the dilation operator $\hat{D}f(x) = \sqrt{2}f(2x)$. A basis function at resolution $k$ and position $m$ is defined as $s^k_m(x) := \hat{D}^k\hat{T}^m s(x) = 2^{k/2}s(2^k x - m)$.

The set of scaling functions at resolution $k$ spans the subspace $\mathcal{H}^k$. The exact relation between resolutions allows $\mathcal{H}^k$ to be embedded in the finer resolution space, $\mathcal{H}^k \subset \mathcal{H}^{k+1}$, creating a nested structure where $\mathcal{L}^2(\mathbb{R})=\lim_{k\to \infty}\mathcal{H}^k$. The relation between scales is governed by the filter coefficients:
\begin{equation}
s^{k}_n(x) = \sum_{m} h_{m-2n} s^{k+1}_m(x).
\end{equation}

The orthogonal complement to $\mathcal{H}^k$ in $\mathcal{H}^{k+1}$ is the wavelet subspace $\mathcal{W}^k$, such that $\mathcal{H}^{k+1}=\mathcal{H}^k\oplus \mathcal{W}^k$. The mother wavelet $w(x)$ is constructed via the quadrature mirror filter relation:
\begin{equation}
w(x) = \sum_{l=0}^{2K-1} g_l \sqrt{2} s(2x - l), \quad \text{where } g_l = (-1)^l h_{2K-1-l}.
\end{equation}
Any function $f(x) \in \mathcal{L}^2(\mathbb{R})$ can be decomposed into a coarse scaling approximation and a sum of wavelet details:
\begin{equation}
f(x) = \sum_{n} f^{s,k}_n s^k_n(x) + \sum_{l=k}^{\infty} \sum_{n} f^{w,l}_n w^l_n(x),
\end{equation}
or, it can also be expanded using only the scaling functions:
\begin{eqnarray}
\label{eq:expansion_scaling_functions}
    f(x) = \lim_{k\rightarrow \infty}\sum_{n} f^{s,k}_n s^k_n(x)
\end{eqnarray}
In this work, we use Daubechies-$3$ ($K=3$), with coefficients listed in Table \ref{tab:h_coeffs}.
\begin{table}[htbp!]
\begin{center}
\setlength{\tabcolsep}{1.0pc}
\newlength{\digitwidth} \settowidth{\digitwidth}{\rm 0}
\catcode`?=\active \def?{\kern\digitwidth}
\caption{$h$ coefficients of Daubechies wavelet for $K=3$.}
\label{tab:h_coeffs}
\vspace{1mm}
\begin{tabular}{c | c }
\specialrule{.15em}{.0em}{.15em}
\hline
$h_0$ & $\dfrac{1+\sqrt{10}+\sqrt{5}+2\sqrt{10}}{16\sqrt{2}}$ \\
$h_1$ & $\dfrac{5+\sqrt{10}+3\sqrt{5}+2\sqrt{10}}{16\sqrt{2}}$ \\ 
$h_2$ & $\dfrac{10-2\sqrt{10}+2\sqrt{5}+2\sqrt{10}}{16\sqrt{2}}$ \\ 
$h_3$ & $\dfrac{10-2\sqrt{10}-2\sqrt{5}+2\sqrt{10}}{16\sqrt{2}}$ \\
$h_4$ & $\dfrac{5+\sqrt{10}-3\sqrt{5}+2\sqrt{10}}{16\sqrt{2}}$ \\ 
$h_5$ & $\dfrac{1+\sqrt{10}-\sqrt{5}+2\sqrt{10}}{16\sqrt{2}}$ \\ 
\hline
\specialrule{.15em}{.15em}{.0em}
\end{tabular}
\end{center}
\end{table}

\begin{figure}[htbp]
\centering
\includegraphics[scale=0.52]{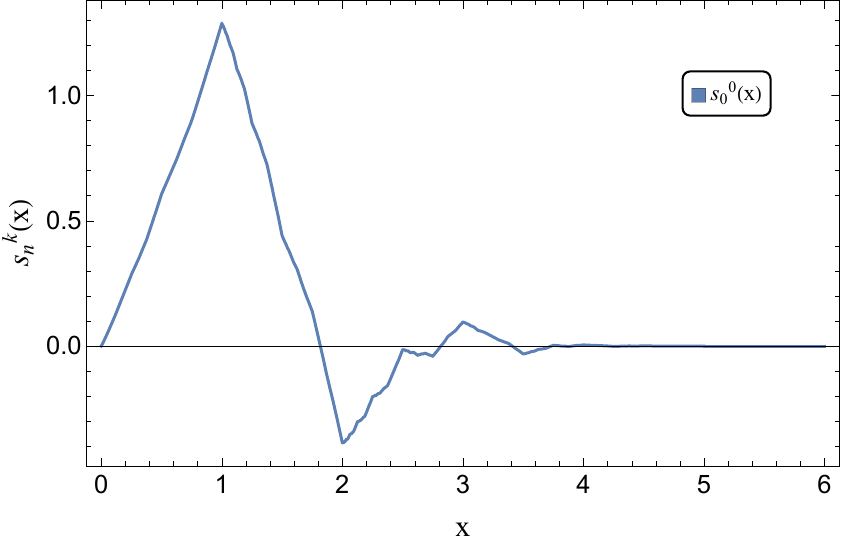}
\includegraphics[scale=0.53]{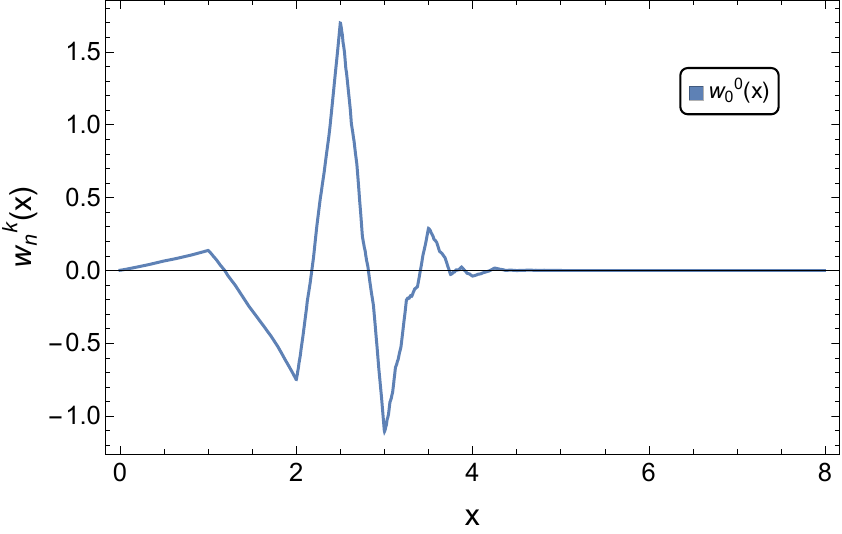}
\caption{The Daubechies-$3$ scaling functions (left) and wavelet functions (right) for resolution $0$ and translation $0$.}
\label{fig:wavelet_basis}
\end{figure}

\section{Wavelet-based Fock Space}
\label{sec:formalism}

In this section, we construct a wavelet-based Fock space for the $1+1$ dimensional free scalar field theory. We derive the Hamiltonian matrix elements in this basis and demonstrate that the wavelet formulation preserves locality through the structure of the hopping amplitudes.

\subsection{Scalar Field Hamiltonian}
The normal-ordered Hamiltonian for a $1+1$ dimensional scalar field $\phi$ with unit bare mass is
\begin{equation}
\label{eq:scalar_field_Hamiltonian_momentum}
\mathrm{H}_0 = \int dp \, E_p \, a^{\dagger}(p)a(p),
\end{equation}
where $E_p = \sqrt{p^2+1}$ and the operators satisfy standard commutation relations. We expand the momentum-space creation and annihilation operators in the scaling basis defined in Eq.(~\ref{eq:expansion_scaling_functions}). 
\begin{eqnarray}
\label{eq:creation_operator_scaling_mode_expansion}
a^{\dagger}(p)&=&\lim_{k\to \infty}\sum_{n=-\infty}^{\infty}a^{s,k\dagger}_n s^k_n(p),\\
\label{eq:annihilation_operator_scaling_mode_expansion}
a(p)&=&\lim_{k\to \infty}\sum_{n=-\infty}^{\infty}a^{s,k}_n s^k_n(p).
\end{eqnarray}
Using the orthogonality of the basis, we invert the expansion to define scaling-mode operators:
\begin{equation}
a^{s,k\dagger}_n = \int dp \, a^{\dagger}(p) s^k_n(p), \quad a^{s,k}_n = \int dp \, a(p) s^k_n(p).
\end{equation}
Physically, $a^{s,k\dagger}_n$ creates a state smeared over a momentum range of width $\approx (2K-1)/2^k$ centered at the translation index $n$. Substituting these into Eq.~(\ref{eq:scalar_field_Hamiltonian_momentum}) yields the following discretized form of the Hamiltonian:
\begin{equation}
\label{eq:the_Hamiltonian_Daubechies_basis}
\mathrm{H}_0 = \lim_{k\to \infty} \sum_{m,n} E^k_{ss,mn} \, a^{s,k\dagger}_m a^{s,k}_n,
\end{equation}
where the hopping matrix elements are given by the overlap integral
\begin{equation}
\label{eq:hopping_strength}
E^k_{ss,mn} = \int dp \sqrt{p^2+1} \, s^k_m(p) s^k_n(p).
\end{equation}
The matrix $E^k_{ss,mn}$ represents the amplitude for a mode to hop between momentum indices $n$ and $m$. Due to the compact support of the Daubechies scaling functions, this matrix is band diagonal; $E^k_{ss,mn}$ vanishes rapidly as $|m-n|$ increases, ensuring the theory remains local in the wavelet representation.

\subsection{Fock Space Construction}
We define a generic multi-particle Fock state $|\Psi\rangle$ by the occupation numbers $\{N_p\}$ of scaling modes at resolution $k$ and position $p$:
\begin{equation}
\label{eq:scaling_mode_fock_basis}
|\Psi\rangle = | \dots, N_p, \dots \rangle,
\end{equation}
where $N_p$ denotes the number of quanta in mode $p$. The operators act typically as $a^{s,k\dagger}_n | \dots, N_n, \dots \rangle = \sqrt{N_n+1} | \dots, N_n+1, \dots \rangle$.
The Hamiltonian matrix elements between two Fock states $|\Psi_1\rangle = \{N^{(1)}_p\}$ and $|\Psi_2\rangle = \{N^{(2)}_p\}$ are non-zero only if particle number is conserved. The matrix element separates into diagonal (number operator) and off-diagonal (hopping) terms:
\begin{equation}
\label{eq:free_hamiltonian_matrix_element}
\langle \Psi_1 | \mathrm{H}_0 | \Psi_2 \rangle = 
\sum_{p,q} E^k_{ss,pq} \bra{\Psi_1} a^{s,k\dagger}_p a^{s,k}_q \ket{\Psi_2}.
\end{equation}
Explicitly, this yields:
\begin{equation}
\mathcal{H}_{1,2} = 
\begin{cases} 
\sum_p E^k_{ss,pp} N^{(2)}_p & \text{if } |\Psi_1\rangle = |\Psi_2\rangle \\
E^k_{ss,pq} \sqrt{(N^{(2)}_p+1) N^{(2)}_q} & \text{if } |\Psi_1\rangle = a^\dagger_p a_q |\Psi_2\rangle \\
0 & \text{otherwise}.
\end{cases}
\end{equation}
This discrete Hamiltonian matrix can be truncated and diagonalized to obtain the energy spectrum at finite resolution $k$.

\subsection{Hamiltonian Eigenvalues}
\label{subsec:eigenvalues}

We diagonalize the Hamiltonian by imposing physical truncations to obtain a finite-dimensional matrix. The truncation scheme involves three constraints:
\begin{enumerate}
    \item \textbf{Momentum Cutoff:} The momentum space volume is restricted by limiting the translation indices of the scaling modes.
    \item \textbf{Energy Cutoff:} We restrict the many-body basis states to those where the total average energy does not exceed a cutoff value $\Lambda$ (set to $10$ in dimensionless units for this simulation).
    \item \textbf{Volume Cutoff:} The resolution $k$ effectively acts as a real-space volume cutoff determined by the support of the basis functions.
\end{enumerate}
We employ periodic boundary conditions for the computation. The eigenvalues obtained by diagonalizing the truncated Hamiltonian for the zero-momentum sector are presented in Table \ref{tab:eigenvalues}.

\begin{table}[htbp]
\centering
\caption{Comparison of exact vs. computed zero-momentum eigenvalues for 1- to 4-particle ground states across increasing resolutions ($k$).}
\label{tab:eigenvalues}
\setlength{\tabcolsep}{1.2pc}
\begin{tabular}{c | c | c | c | c}
\specialrule{.15em}{.0em}{.15em}
\hline
State & Exact & $k=0$ & $k=1$ & $k=2$ \\
\hline
$1p$ & $1$ & $1.04363$ & $1.01152$ & $1.00408$ \\
$2p$ & $2$ & $2.08726$ & $2.02304$ & $2.00815$ \\
$3p$ & $3$ & $3.13093$ & $3.03457$ & $3.01223$ \\
$4p$ & $4$ & $4.17466$ & $4.04613$ & $4.01630$ \\
\hline
\specialrule{.15em}{.0em}{.15em}
\end{tabular}
\end{table}

As shown in Table \ref{tab:eigenvalues}, the computed eigenvalues converge rapidly toward the exact values as the resolution $k$ increases. This convergence in the free theory validates the efficacy of the wavelet discretization and motivates the application of this formalism to interacting theories, specifically $\phi^4$-theory, which we discuss in the following section.

\section{The $\phi^4$ Hamiltonian in Wavelet Basis}
\label{sec:phi4_hamiltonian}

In this section, we extend the formalism to the interacting $1+1$ dimensional $\phi^4$ theory. We derive the Hamiltonian matrix elements in the wavelet Fock space, compute the energy spectrum for varying coupling strengths $\lambda$, and analyze the critical behavior associated with $\mathbb{Z}_2$ symmetry breaking.

\subsection{Interaction Hamiltonian Construction}
The normal-ordered interaction Hamiltonian is given by:
\begin{equation}
\mathrm{H_I} = \frac{\lambda}{4!} \int dx :\!\phi^4(x)\!: \quad (\lambda > 0).
\end{equation}
By expanding the field operators in the scaling basis, the interaction term takes the discretized form:
\begin{equation}
\label{eq:interacting_hamiltonian_scaling_particl_mode}
\mathrm{H_I} = \frac{\lambda}{4!(4\pi)^2} \sum_{\{q_i\}} \Gamma^k_{q_1 q_2 q_3 q_4} \, : \prod_{i=1}^4 \hat{\mathcal{A}}_{q_i} :,
\end{equation}
where $\hat{\mathcal{A}}_{q_i}$ represents either a creation ($a^{s,k\dagger}_{q_i}$) or annihilation ($a^{s,k}_{q_i}$) operator. The vertex factor $\Gamma^k$ is the momentum-space overlap integral of four scaling functions:
\begin{equation}
\label{eq:phi4_hopping_term}
\Gamma^k_{\{q_i\}} = \int \frac{ \prod_{j=1}^4 dp_j \, s^k_{q_j}(p_j) }{\sqrt{\prod_{j=1}^4 E_j}} \, \delta(\sum p_j),
\end{equation}
where $E_j = \sqrt{p_j^2+m^2}$. Efficient evaluation of these integrals is detailed in \cite{basak2026hamiltonianformulation11dimensionalphi4}.

The matrix elements of $\mathrm{H_I}$ between two Fock states $|\Psi_1\rangle$ and $|\Psi_2\rangle$ are calculated similarly to the free case. The non-vanishing terms correspond to the standard normal-ordered operator combinations ($a^\dagger a^\dagger a^\dagger a^\dagger$, $a^\dagger a^\dagger a^\dagger a$, etc.):
\begin{align}
\langle \Psi_1 | \mathrm{H_I} | \Psi_2 \rangle &= \frac{\lambda}{4!(4\pi)^2} \sum_{\{q_i\}} \Gamma^k_{\{q_i\}} \Big[ 
\mathcal{A}^k - 4\mathcal{B}^k + 6\mathcal{C}^k - 4\mathcal{D}^k + \mathcal{E}^k 
\Big].
\end{align}
Here, the terms $\mathcal{A}^k$ through $\mathcal{E}^k$ represent the matrix elements of the operator sequences $a a a a$, $a^\dagger a a a$, $a^\dagger a^\dagger a a$, $a^\dagger a^\dagger a^\dagger a$, and $a^\dagger a^\dagger a^\dagger a^\dagger$, respectively. For details calculation see \cite{basak2026hamiltonianformulation11dimensionalphi4}.

\subsection{Spectrum and Criticality}
We diagonalize the total Hamiltonian $H = H_0 + H_I$ for resolutions $k=0$ and $k=1$ across a range of coupling constants $\lambda$. The low-lying energy eigenvalues are listed in Table \ref{tab:eigenvalues_phi_4}.

\begin{table*}[htbp]
\caption{Energy eigenvalues of $1+1$ dimensional $\phi^4$ theory for resolutions $k=0$ and $k=1$ with increasing coupling $\lambda$.}
\label{tab:eigenvalues_phi_4}
\centering
\setlength{\tabcolsep}{0.6pc}
\begin{tabular}{c | c c c c | c c c c}
\specialrule{.15em}{.0em}{.15em}
\hline
\multirow{2}{*}{$\lambda$} & \multicolumn{4}{c|}{$k=0$} & \multicolumn{4}{c}{$k=1$} \\
\cline{2-9}
& $E_0$ & $E_1$ & $E_2$ & $E_3$ & $E_0$ & $E_1$ & $E_2$ & $E_3$ \\
\hline
0   & $0.000$ & $1.044$ & $1.419$ & $1.451$ & $0.000$ & $1.011$ & $1.122$ & $1.133$ \\
20  & $-0.002$ & $1.034$ & $1.411$ & $1.444$ & $-0.005$ & $0.999$ & $1.111$ & $1.122$ \\
40  & $-0.009$ & $1.006$ & $1.385$ & $1.425$ & $-0.019$ & $0.959$ & $1.071$ & $1.092$ \\
60  & $-0.022$ & $0.950$ & $1.318$ & $1.395$ & $-0.047$ & $0.851$ & $0.952$ & $1.036$ \\
70  & $-0.032$ & $0.900$ & $1.234$ & $1.375$ & $-0.073$ & $0.583$ & $0.626$ & $0.706$ \\
80  & $-0.045$ & $0.801$ & $1.012$ & $1.271$ & $-0.280$ & $-0.128$ & $-0.003$ & $0.033$ \\
90  & $-0.066$ & $0.471$ & $0.636$ & $0.794$ & $-1.133$ & $-0.887$ & $-0.781$ & $-0.597$ \\
100 & $-0.127$ & $-0.120$ & $0.170$ & $0.241$ & $-2.036$ & $-1.743$ & $-1.603$ & $-1.381$ \\
120 & $-1.468$ & $-1.089$ & $-1.026$ & $-0.800$ & $-3.890$ & $-3.511$ & $-3.305$ & $-3.010$ \\
140 & $-2.872$ & $-2.422$ & $-2.320$ & $-1.970$ & $-5.776$ & $-5.314$ & $-5.047$ & $-4.680$ \\
\hline
\specialrule{.15em}{.0em}{.15em}
\end{tabular}
\end{table*}

\begin{figure}[htbp]
\centering
\includegraphics[scale=0.49]{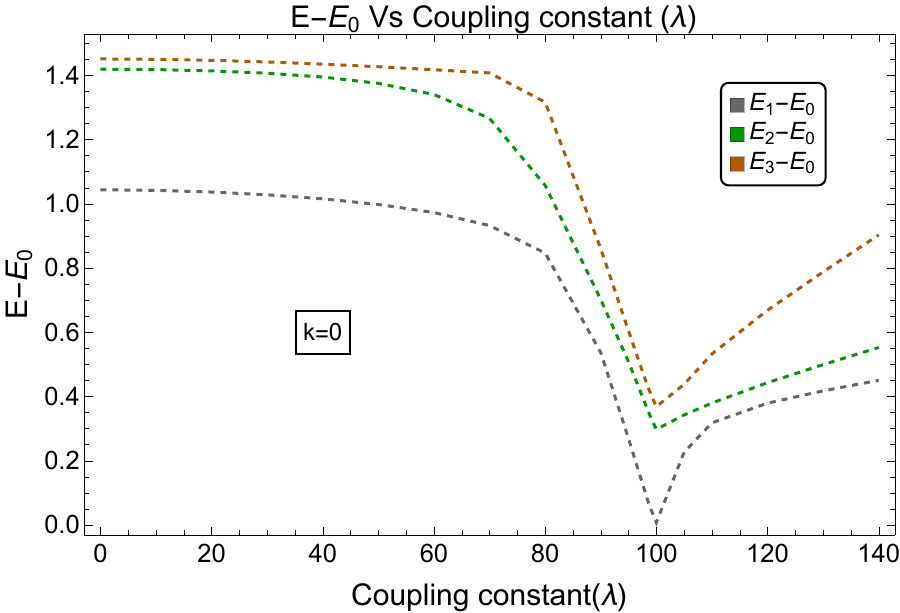} 
\includegraphics[scale=0.49]{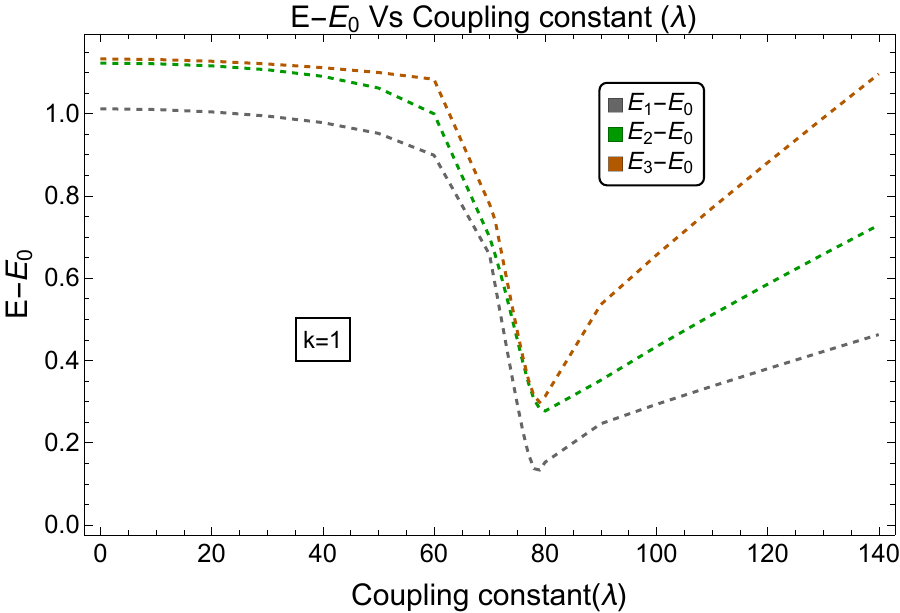} 
\caption{The energy gap ($E_1 - E_0$) as a function of coupling $\lambda$. The sharp drop indicates the onset of the $\mathbb{Z}_2$ symmetry breaking phase transition.}
\label{fig:energy_gap}
\end{figure}

Two key features are observed:
\begin{enumerate}
    \item \textbf{Vacuum Energy Divergence:} The ground state energy $E_0$ becomes increasingly negative with $\lambda$ (see Table \ref{tab:eigenvalues_phi_4}). This is a known artifact of defining the Hamiltonian relative to the free-theory normal-ordered vacuum ($E_{vac}=0$). The attractive interaction lowers the energy below this reference point.
    
    \item \textbf{Symmetry Breaking:} We analyze the gap between the ground state and the first excited state ($E_1 - E_0$). As shown in Fig. \ref{fig:energy_gap}, the gap closes near a critical coupling $\lambda_c$, signaling the onset of degeneracy associated with $\mathbb{Z}_2$ symmetry breaking. The subsequent lifting of this degeneracy is an artifact of the finite resolution and the asymmetry of the Daubechies basis functions.
\end{enumerate}

We estimate the critical point to be $\lambda_c \approx 100$ ($k=0$) and $\lambda_c \approx 79$ ($k=1$). In terms of the critical coupling $g_c = \lambda_c/24$, this corresponds to $g_c \approx 4.17$ and $g_c \approx 3.29$, respectively. Increasing the resolution shifts the critical value toward the established literature value of $g_c \sim 2.8$  \cite{PhysRevD.79.056008, Bosetti_2015, PhysRevD.99.034508, PhysRevD.88.085030, PhysRevD.106.L071501, 10.1016/j.physletb.2015.11.015, 10.1007/JHEP08(2018)148, PhysRevD.93.065014, 10.1007/JHEP05(2019)184, PhysRevD.96.065024, demiray2025systematicimprovementhamiltoniantruncation, PhysRevResearch.2.033278, PhysRevD.109.045016}. This trend indicates that higher resolutions will further improve the accuracy of the critical point estimation.

\section{Conclusion and Outlook}
\label{sec:conclusion}

In this work, we developed a nonperturbative Hamiltonian formulation of the $1+1$ dimensional $\phi^4$ theory using a momentum-space Daubechies wavelet basis, realizing Wilson’s original vision of a "wave-packet" expansion. By decomposing field operators into wavelet modes characterized by resolution and translation indices, we constructed a systematic truncation scheme that serves as a robust alternative to standard Fourier discretization.

Our results highlight two distinct advantages of this framework. First, the compact support of the Daubechies wavelets results in a free Hamiltonian with finite-range hopping (locality preservation). Second, in the interacting $\phi^4$ case, the matrix representation exhibits high compressibility, as the dominant contributions arise from a limited number of degrees of freedom.

Crucially, we demonstrated that the quantum critical point can be reliably extracted even at coarse resolutions. Our estimated critical couplings converge toward established literature values as the resolution increases, validating the consistency of the method.

Looking ahead, we identify several promising directions for this formalism:
\begin{enumerate}
    \item \textbf{Optimization:} Future work will implement projection techniques to the zero-momentum sector to reduce computational costs, enabling access to higher resolutions and more precise spectra.
    \item \textbf{Higher Dimensions \& Gauge Theories:} The multiresolution structure and inherent locality of the basis make it well-suited for extension to higher-dimensional systems and potentially gauge theories or QCD-like models.
    \item \textbf{Position-Space Formulations:} We are currently developing a complementary position-space wavelet Hamiltonian approach, which will be reported in subsequent studies.
\end{enumerate}

In summary, this work establishes the momentum-space wavelet basis as a scalable and effective tool for Hamiltonian truncation in quantum field theory.


\acknowledgments
This work is supported by the Department of Atomic Energy, Government of India, under Project Identification Number RTI 4002. Computations were carried out on the computing clusters at the Department of Theoretical Physics, TIFR, Mumbai. I would like to thank Prof. Nilmani Mathur, Debsubhra Chakraborty and Prof. Raghunath Ratabole for their useful comments. I would also like to thank  Ajay Salve and Kapil Ghadiali for computational support.


\bibliographystyle{JHEP}
\bibliography{References}
\end{document}